\documentclass[]{aa}   

\usepackage{graphicx}
\usepackage{txfonts}
\usepackage{xcolor}
\usepackage{appendix}
\usepackage{caption}
\usepackage{subcaption}
\usepackage{ulem}

\newcommand{\WASP}{WASP-77Ab}
\newcommand{\metallicity}{[(C+O)/H]}

\begin{document}

   \title{Retrieving planet formation parameters of WASP-77Ab using SimAb}

   \author{N. Khorshid
          \inst{1,2,3}
          \and
          M. Min
          \inst{1,2}
          \and
          J.M. D\'esert
          \inst{1}
          }

    \institute{Anton Pannekoek Institute for Astronomy, University of Amsterdam,
              Science Park 904, NL-1098XH Amsterdam\\
         \and
             SRON Leiden,
             Niels Bohrweg 4,
             2333 CA Leiden, the Netherlands\\
        \and
            Department of Space, Earth and Environment, Chalmers University of Technology, Gothenburg SE-412 96, Sweden\\
             \email{niloofar.khorshid@chalmers.se}\\
             }

   \date{Received Month 00, 2021; accepted Maonth 00, 2021}

  \abstract
   {The atmospheric compositions of planets offer a unique view into their respective formation processes. State-of-the-art observatories and techniques are finally able to provide high-precision data on atmospheric composition that can be used to constrain planet formation.}
   {In this context, we focus on the formation of \WASP ~based on previous observations of its atmosphere, which have provided precise C/O and metallicity measurements.}
   {We use the SimAb planet formation simulation to model the formation of~\WASP . We assume two compositions for the disk \WASP ~was formed within: one of a solar composition and one that represents the composition of WASP-77A. In addition, we considered two different scenarios regarding the migration of the planet and we study the possible planet formation paths that reproduce the composition of \WASP. }
   {This work shows that the planet is expected to have formed in a disk where not many planetesimals could be accreted. Moreover, we demonstrate that the most likely migration scenario is disk-free migration, whereby the planet initiates its Type II migration within the CO ice line and ends it beyond the water ice line.}


   \maketitle

\section{Introduction}
\label{sec: introduction}

    {Planet formation studies are aimed at linking the compositions of fully formed planets and their formation histories \citep{Mordasini_2016,Cridland_2019, pacetti_2022}. In particular, the abundances of carbon and oxygen, as well as their ratios, are considered key with regard to both planet formation and planetary atmospheres among their respective communities \citep{Madhusudhan_2014,Pelletier_2021,Dash_2022,Yan_2022}. Planet formation studies show a link between the C/O ratio and where a planet was formed within the disk \citep{Oberg_2011, Notsu_2020}. These studies show that such links may not be straightforward due to the many complex processes taking place during planet formation as well as within planetary atmospheres. Any links may be dependent on the assumptions involved in the models used for these studies \citep{Molliere_2020}.
    
    A recent observation of \WASP ~by \citet{Line_2021} has provided high-precision information on the carbon and the oxygen abundances of its atmosphere. This observation represents an opportunity to tackle the question of the formation history of \WASP .
    
    \citet{Line_2021} suggest that WASP-77 A b accreted its envelope interior to its parent protoplanetary disk's H$_2$O ice line from carbon-depleted gas with little subsequent planetesimal accretion or core erosion. However, the non-solar abundance ratios of WASP-77 A reported by \citet{Reggiani_2022} have changed the initial interpretation of the formation of \WASP . \citet{Reggiani_2022} finds that the super-stellar C/O ratio of the planet implies formation outside its parent protoplanetary disk's H$_2$O ice line.

     In this study, we use a Bayesian retrieval formation to couple the observation in \citet{Line_2021} to the formation parameters introduced in \citet{khorshid_2021} (from now on referred to as Paper I). We use the formation code Simulating Abundances (SimAb), which is a fast and flexible planet formation model. SimAb connects the atmospheric abundances of gas giants to their formation path including their formation location, and how much solid material was accreted during their formation.

    In Section \ref{sec1: int_wasp}, we introduce the WASP-77 planetary system. The method is outlined in Section \ref{sec: method}, where we explain the formation model and scenarios used in this study and how the retrieval was set up. In Section \ref{sec: results}, we report the results of this study and the Bayesian evidence of the different scenarios that were included in this study. These results are discussed in Section \ref{sec: discuss}, where we discuss the most likely planet formation scenario for \WASP ~and compare it to the results of previous studies on its formation. Finally, we present our conclusions  in Section \ref{sec: conclusion}.

   \section{WASP-77Ab} 
   \label{sec1: int_wasp}
     WASP-77 Ab is a hot Jupiter orbiting around the brighter star of a visual binary. The planet has a radius of $1.21\pm0.02R_J$ and a mass of $1.76\pm0.06M_J$. Its orbit has a period of $1.36$ days at a distance of $0.024$AU \citep{Maxted_2013}. A recent study by \citet{Mansfield_2022} of the atmosphere of \WASP ~suggests an equilibrium temperature of 2000K and little to no cloud coverage on the day side. Observations by \citet{Salz_2015} in the {X-ray} suggest a rapid mass loss for the planet. The same study suggests the presence of smaller planetary companions based on the periodic analysis of the star which has not been observed yet.
     
     The host star is a G8 type star with a temperature of around 5600K \citep{Maxted_2013}. The reported age of the star varies between 7.57 and 1.35 Gyr depending on the model used \citep{Maxted_2015}. \citet{Reggiani_2022} show that despite the solar-like metallicity $[Fe/H] = -0.01$, WASP-77A is enhanced both in carbon and oxygen abundances, which results in [(C+O)/H] = 0.33 relative to the solar value in logarithmic scale. \citet{Evans_2018} reports a bimodal distribution for the eccentricity of the system, 0.5 or 0.95, which is a characteristic of wide binaries.
 \section{Method}
\label{sec: method}
    \subsection{Observed composition of WASP-77Ab}
    \label{sec1: met-W77}
    In this study, we use the posterior distribution retrieved by \citet{Line_2021} for the elemental abundances of oxygen and carbon. These data are based on the observations of the secondary eclipse of \WASP ~on December 14, 2020, using the Immersion GRating INfrared
Spectrometer (IGRINS) at Gemini South. To understand the C/O ratio and \metallicity ~of \WASP,~these authors used pymultinest \citep{pymultinest}. They constrained some of the main molecules in the atmosphere (e.g., H$_2$O, CO, CH$_4$, H$_2$S, NH$_3$, and HCN), along with the vertical temperature structure, planetary orbit, and system velocity. 
    
    We use the retrieved abundances reported by \citet{Line_2021} to calculate the C/O ratio and the \metallicity ~of \WASP, ~as described therein. The CO abundance is used to calculate the carbon abundance and the CO and H$_2$O abundances are used to calculate the oxygen abundance. The $H_2$ molecule is assumed to form $83.1$ percent of the atmosphere. We used the solar abundances based on the work of \citet{Asplund_2009}.

    \subsection{Formation model}
    \label{sec1: met_SimAb}

    In this study, we use SimAb \citep{khorshid_2021}, a planet formation simulation code that predicts the atmospheric abundances of gas giant planets based on their formation scenarios. The code calculates the atmospheric elemental abundances of the planet based on the initial orbital distance from where the planet begins accreting its atmosphere ($R_{start}$), along with its efficiency in accreting planetesimals ($f_{p}$) and the percentage of the solid phase that can be coupled to the gas ($f_{dust}$). It adjusts the disk viscosity (through the $\alpha$ parameter) so that the planet reaches the given orbital distance by the time its mass reaches the final mass. The abundances of the atoms are then calculated from the accreted mass of the gas phase and the solid phase from the different parts of the disk with different atomic fractions. Paper I shows that the initial core mass of the planet does not influence the C/O ratio or the metallicity of the atmosphere. Hence, in this work we assume a fixed mass of 20 Earth masses for the core of the planet. For further details, we refer to Paper I.

    \subsubsection{Soot line}
    \label{sec2: sootline}
    In addition to the three parameters introduced in Paper I, which impact the atmospheric compositions of the gas giants, we introduce a new parameter that is the amount of carbon ($f_{carbon}$) locked up in solids beyond the so-called soot line. The soot line is located in the disk at a radius where the carbonaceous materials are exposed to very high temperatures and thus destroyed by oxidation processes \citep{Hoff_2020}. For locations beyond the soot line, the carbonaceous materials are not affected and are still available in the disk. In this study, we choose a conservative temperature of 800K for the location of the soot line \citep{KRESS_2010}. The carbon fraction in the solid phase at temperatures lower than this temperature is set to $f_{carbon}$. 
    
    In Paper I, we assumed 10\%\ of carbon goes into the solid phase to match the observed carbon abundances in the meteorite materials reported by \citet{Asplund_2009}. In the current study, we allow this percentage to vary, which impacts the C/O ratio in the solid phase and the gas phase at different ice lines. This allows the formation of planets with C/O ratio and metallicities that was not initially predicted in Paper I.
    
    To include a soot line, we vary the $f_{carbon}$ between zero and one when calculating the atomic abundances in the solid phase and the gas phase in the disk. A $f_{carbon}$ value of 0 means there is no carbon in the solid phase at 800 K, while the other disk abundances are the same as they are in Paper I. A carbon fraction of 1 means that all the carbon is in the solid phase at 800K. For any given carbon fraction, we add a new step at 800K in calculating the atomic abundance in the disk, done using the abundance calculating module explained in Paper I. At 800K, the carbon abundance in the solid phase compared to the gas phase is set to $f_{carbon}$.

    \subsubsection{Migration path}
    \label{sec2: Rend}

    Even though some studies look at the possibility of in situ formation of gas giants \citep{boley_2016,Bailey_2018}, hot Jupiters are mainly assumed to have migrated to their current position \citep{Ida_2004,Alibert_2005,Papaloizou_2007}. The planet's migration can be caused by torques caused by the disk \citep{Lin_1986,Kley_2012} or through dynamical interactions with a third body \citep{Fabrycky_2007}. To include disk-free migration in the formation simulation, we introduce migration endpoint $R_{end}$ and migration distance $R_{migrate}$. These parameters would thus replace $R_{start}$. In this situation, $R_{start}$ is calculated using Eq.~(\ref{eq: distance}):
    
    \begin{equation}
    \label{eq: distance}
        R_{start}=R_{end}+R_{migrate}
    ,\end{equation}
    
    When it comes to $R_{end}$, it can be the same as the current position of the planet that probes scenarios where the planet migrates all the way to its current position through Type II migration (the scenario from Paper I). When $R_{end}$ is set to be a value different from the current position of the planet, the planet migrates to $R_{end}$, while accreting material and reaching its mature mass. Following this, the planet moves to its current position through disk-free migration without accreting any further material. This scenario assumes there is no further planetesimal accretion after the planet migrates to its final position. Considering the low metallicity of the atmosphere of the planet, late accretion of planetesimals is very unlikely. Furthermore, this assumption can be supported by \citet{Guillot_2000} where the authors discuss the fact that planetesimal accretion is more efficient within the first 10 Myr, where there is still gas in the disk. By setting $R_{migrate}$ to values very close to zero, we can study in situ formation. However, due to the computational restrictions of the model, $R_{migrate}$, can never be set to zero; therefore, in this study, we assume a lower limit of 0.001AU.
    
    \subsubsection{\WASP ~formation scenarios}
    \label{sec2: formation scenarios}
    To simulate the formation of \WASP ,~we set the final mass to 1.76 $M_{Jupiter}$ and the final orbital distance to $0.024$ AU. We assume the mass and the luminosity of the star is the same as they are for the Sun. In this study, we adjusted the disk abundance to represent the composition of WASP-77A. To do so, we chose the abundance of the elements to be the same as the reported values in \citet{Asplund_2009}, then we adjust the carbon and oxygen abundance to $3.12\times10^{-4} \cdot[H]$ and $7.66\times10^{-4}\cdot[H],$ respectively, to match the values reported in \citet{Reggiani_2022}, based on their 3D non-LTE assumption. We assumed two different formation scenarios for a planet with similar C/O ratio and \metallicity , as reported in \citet{Line_2021}. We describe these scenarios below. 
    
    \textbf{Scenario 1}: This scenario assumes the planet has migrated to its current orbital distance through Type II migration, while accreting its atmosphere during the process. In this scenario, we set the final orbital distance, $R_{end}$, equal to the current orbital distance of \WASP . In this case, retrieving the initial orbital distance is equivalent to retrieving the migration distance. In the first scenario $R_{start}$, $f_{p}$, $f_{dust}$, and $f_{carbon}$ are the parameters we retrieve.
    
    \textbf{Scenario 2}: Here, we focus on the disk-free migration scenario. We allow all the parameters to vary, including the migration endpoint. The composition of the planet is assumed to not change after the Type II migration is stopped. Figure~\ref{Fig: regions} shows the migration path of these scenarios as well as the different regions with respect to the three main ice lines we include in this study. In this scenario $R_{end}$, $R_m$, $f_{p}$, $f_{dust}$, and $f_{carbon}$ are the retrieving parameters. The locations of the ice lines are given in Table \ref{tab: iceline}

        \begin{figure}[!htbp]
                \centering
                \includegraphics[height=4.5cm]{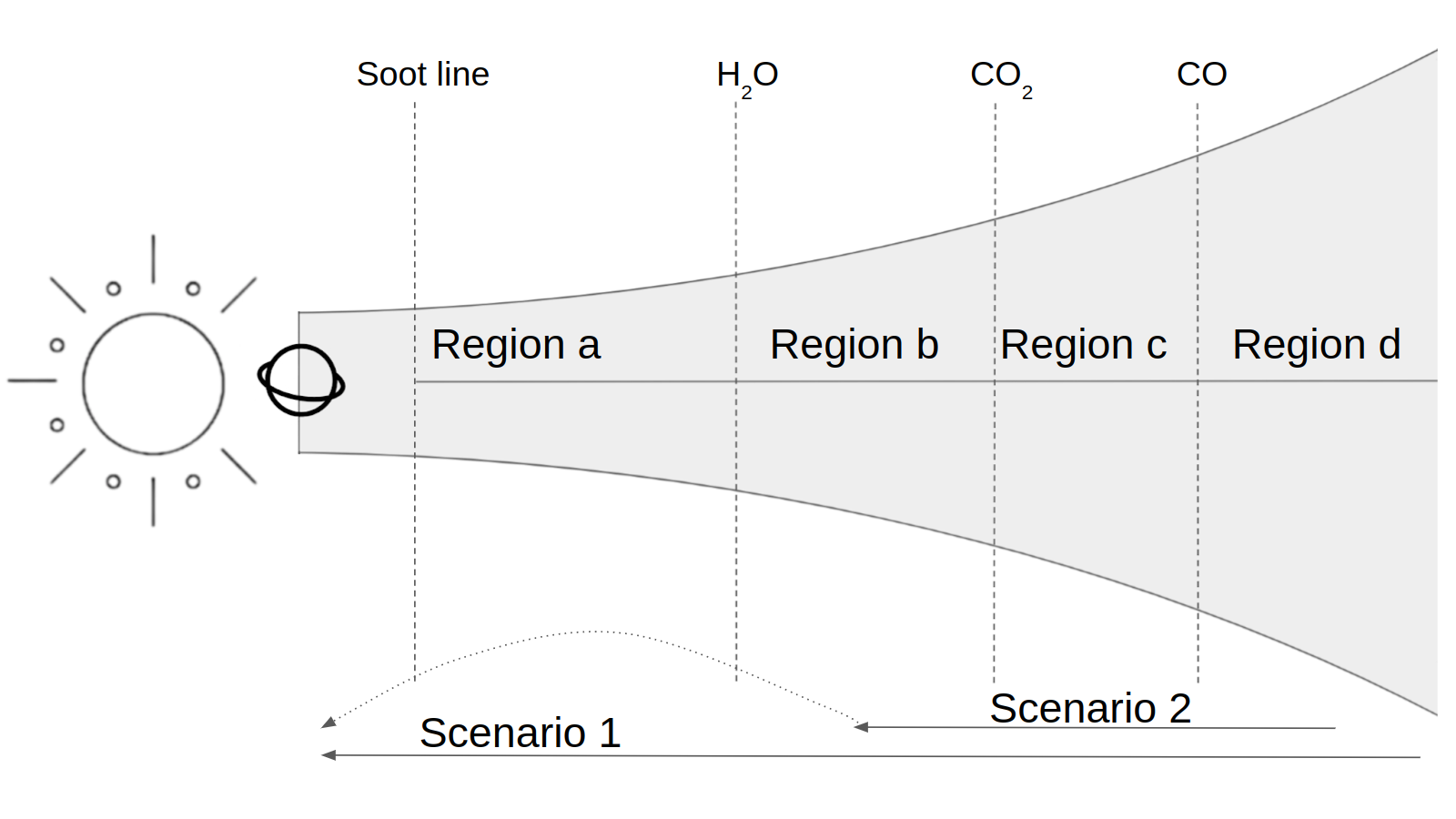}
                \caption{Different regions in the disk with respect to the ice lines. Arrows on the bottom of the disk show a schematic of the migration path for the two scenarios considered in this study. The planet shows the current position of \WASP . The ice line locations as well as the planet location is not representing the actual distance from the host star.}
                \label{Fig: regions}
        \end{figure}
    
    \begin{table}[ht!]
                \begin{center}
                        \begin{tabular}{ |c|c|c| } 
                                \hline
                                Ice line&Temperature (K)&Location (AU)\\
                                \hline
                                   
                                Carbon soot line&800&0.54\\
                            H$_2$O &120&4.5\\
                                CO$_2$ &47&15.00\\
                                CO &20&110.11\\    
                                \hline
                                
                        \end{tabular}
                        \caption{Location and temperature of the three main ice lines and the carbon soot line}
                        \label{tab: iceline}
                \end{center}
                
        \end{table}

    \subsection{Retrieval}
    \label{sec1: met_retrieval}
   
    In order to be able to retrieve the formation parameters described in Sect. (\ref{sec2: formation scenarios}), we use MultiNest \citep{Feroz_2009} to sample the parameter space and find their likelihood of forming planets with a similar composition to what has been observed for \WASP ~using SimAb. We set the live points to be 1000 points and the convergence factor to be 0.05. 
    
    The parameters that are retrieved are presented in Table \ref{tab: parameters1} for the first scenario and Table \ref{tab: parameters2} for the second scenario. A uniform linear prior is assumed for all of these parameters. The reason for choosing a linear sampling for the orbital distances is to make sure that SimAb output is spread across the C/O ratio and \metallicity ~plane as uniformly as possible. Figure~\ref{Fig: run-all} (\ref{Fig: run0} and \ref{Fig: runol}) shows the C/O and \metallicity ~distributions of the planets. This figure also shows that sampling from a linear distribution is more spread throughout the C/O ratio and \metallicity ~plane, while the logarithmic sampling is focused on solar composition. The explanation for this is that the logarithmic sampling focuses the prior very much on the inner regions, within all the major ice lines, and therefore produces mostly planets with Solar \metallicity ~and C/O ratio.
    
    We calculate the likelihood from the atmospheric composition posterior distribution derived by \citet{Line_2021}. In Section \ref{sec1: met-W77}, we explain the steps required to obtain this posterior distribution. The posterior distribution that represents the C/O ratio and \metallicity ~in \citet{Line_2021} shows a correlation between C/O ratio and \metallicity . Therefore, using the posterior distribution is more accurate compared to using the error bars reported for C/O and \metallicity ~in \cite{Line_2021}.
    
    The atmospheric composition posterior distribution consists of discreet points, therefore, a Gaussian smoothing kernel is used, with a width of one-twentieth of the error bars in each dimension. This choice for the width of the smoothing kernel allows for the posterior distribution to be sampled in a way that keeps the characteristics of the distribution while making sure that there are enough overlaps between the samples in the posterior distribution file, so that the posterior distribution is no longer discreet. An oversmoothed posterior distribution would result in a low precision level for the retrieved formation parameters; whereas a posterior distribution that is not adequately smoothed would result in unreliable retrieved values for the formation parameters. To compute the likelihood of point $x_0$ and $y_0$, using the points from the posterior distribution of the atmospheric composition ($x_i$, $y_i$), we use Eq.(\ref{Eq: logliklyhood}): 
    
    \begin{equation}
    L = \frac{\sum\exp^{-1/2((\frac{(x_i - x_0}{\sigma_{x}})^2+(\frac{y_i - y_0}{\sigma_{y}})^2)}}{2\pi \sigma_{x} \sigma_{y}N}
        \label{Eq: logliklyhood}
    .\end{equation}

    In this equation, 'N' is the number of points in the posterior distribution of the atmospheric composition. Then, $\sigma_x = 0.004$ and $\sigma_y= 0.008$ are the widths of each Gaussian in the C/O ratio and \metallicity ~plane, respectively.

    \begin{figure}
     \centering
     \begin{subfigure}[b]{0.5\textwidth}
         \centering
         \includegraphics[height=4.5cm]{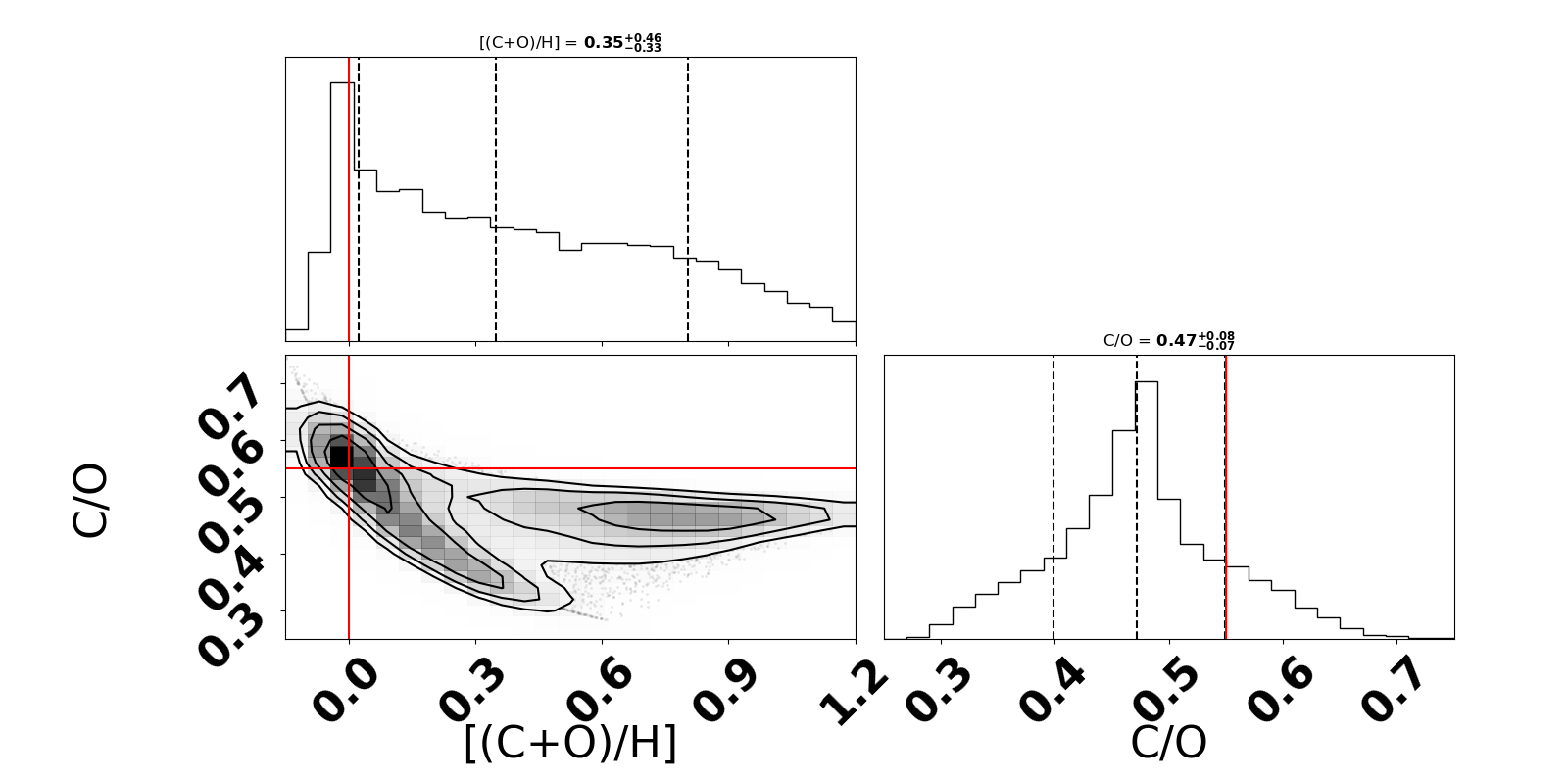}
        \caption{}
        \label{Fig: run0}
     \end{subfigure}
     \hfill
     \begin{subfigure}[b]{0.5\textwidth}
         \centering
        \includegraphics[height=4.5cm]{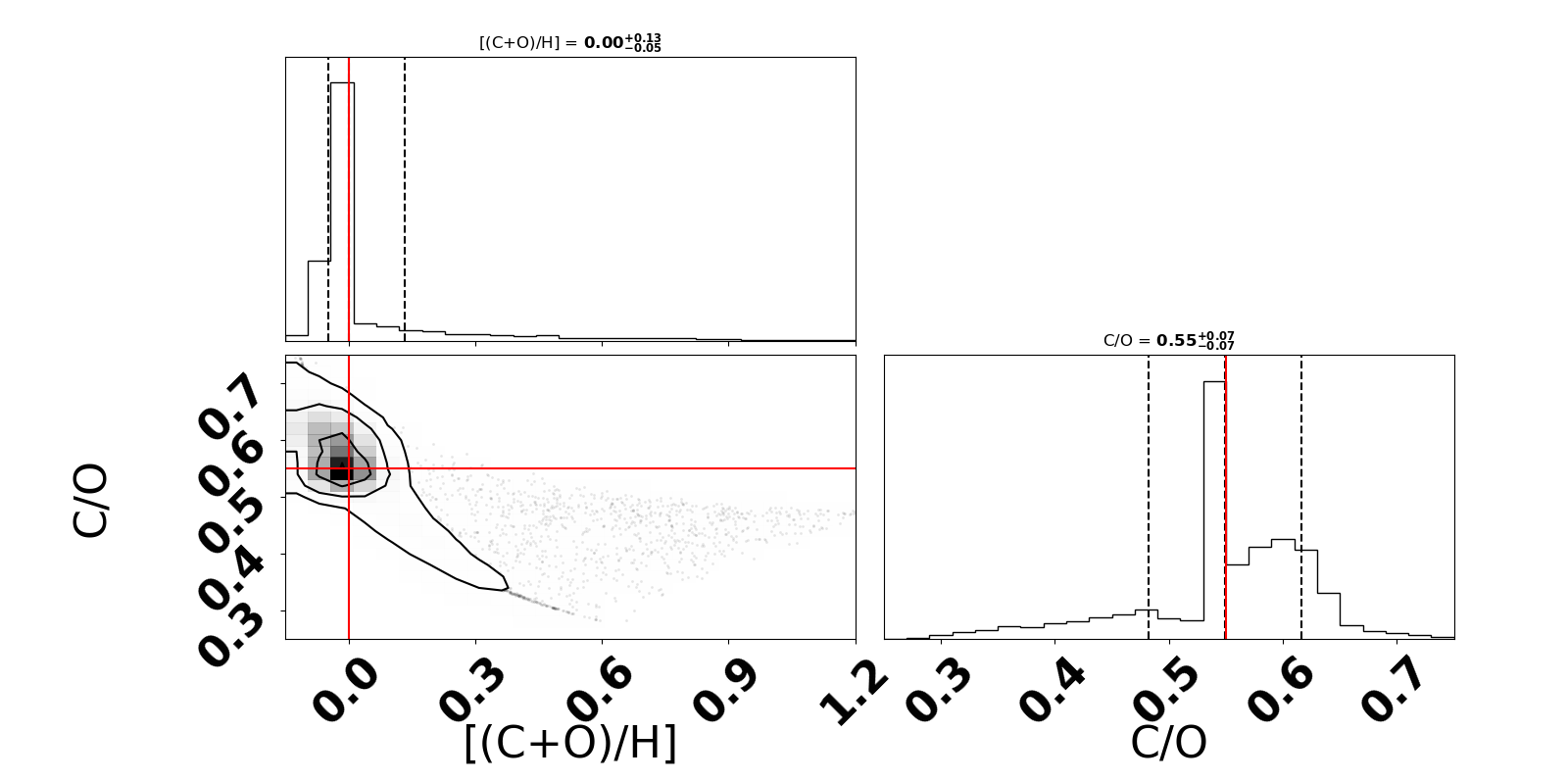}
        \caption{}
        \label{Fig: runol}
     \end{subfigure}
     
     \caption{C/O ratio and the logarithm of \metallicity , compared to solar, distribution of planets with the mass and orbital distance of \WASP . Red line shows the solar C/O ratio and \metallicity . The initial orbital distance is randomly chosen from a uniform distribution in linear (logarithmic) space for the top (bottom) panel from 300 AU to the current position of the planet.}
     \label{Fig: run-all}
\end{figure}

        \begin{table}[ht!]
                \begin{center}
                        \begin{tabular}{ |c|c|c|c| } 
                                \hline
                                Parameter & symbol & range & distribution\\
                                \hline
                                   
                                Migration start point&$R_{start}$&0.024 - 300&uniform\\
                                Planetesimal ratio &$f_{p}$&0-1&uniform\\
                                Dust grain fraction &$f_{dust}$&0-1&uniform\\
                                Carbon fraction &$f_{carbon}$&0-1&uniform\\    
                                \hline
                                
                        \end{tabular}
                        \caption{\textbf{Scenario 1:} Retrieval parameters.}
                        \label{tab: parameters1}
                \end{center}
                
        \end{table}

        \begin{table}[ht!]
                \begin{center}
                        \begin{tabular}{ |c|c|c|c| } 
                                \hline
                                Parameter & symbol & range & distribution\\
                                \hline
                                Migration endpoint&$R_{end}$&0.024-200&uniform\\  
                                Migration distance & $R_{d}$&0-200&uniform\\  
                                Planetesimal ratio &$f_{p}$&0-1&uniform\\
                                Dust grain fraction &$f_{dust}$&0-1&uniform\\
                                Carbon fraction &$f_{carbon}$&0-1&uniform\\    
                                \hline
                                
                        \end{tabular}
                        \caption{\textbf{Scenario 2:} Retrieval parameters.}
                        \label{tab: parameters2}
                \end{center}
                
        \end{table}

\section{Results}
\label{sec: results}

        We used the observed C/O ratio and \metallicity ~of \WASP ~reported by \citet{Line_2021} to retrieve its formation parameters with the SimAb formation simulation. In the following, we present the results of the retrievals and we describe how well the retrieved values reproduce the observed C/O ratio and \metallicity ~of \WASP . The retrievals are done based on the two scenarios that are explained in the method section.
        
        Figure~\ref{Fig: init} shows the C/O ratio and \metallicity ~of the planets formed with the same mass and orbital distance as \WASP ~using SimAb. This figure shows that by assuming the same planet formation scenario as was reported in Paper I (i.e. $f_{carbon}=0, R_{end}=0.024\,$AU), the composition of the simulated planets would not be within three sigma accuracy of the observed composition of \WASP .

        \begin{figure}[!htbp]
                \centering
                \includegraphics[height=4.5cm]{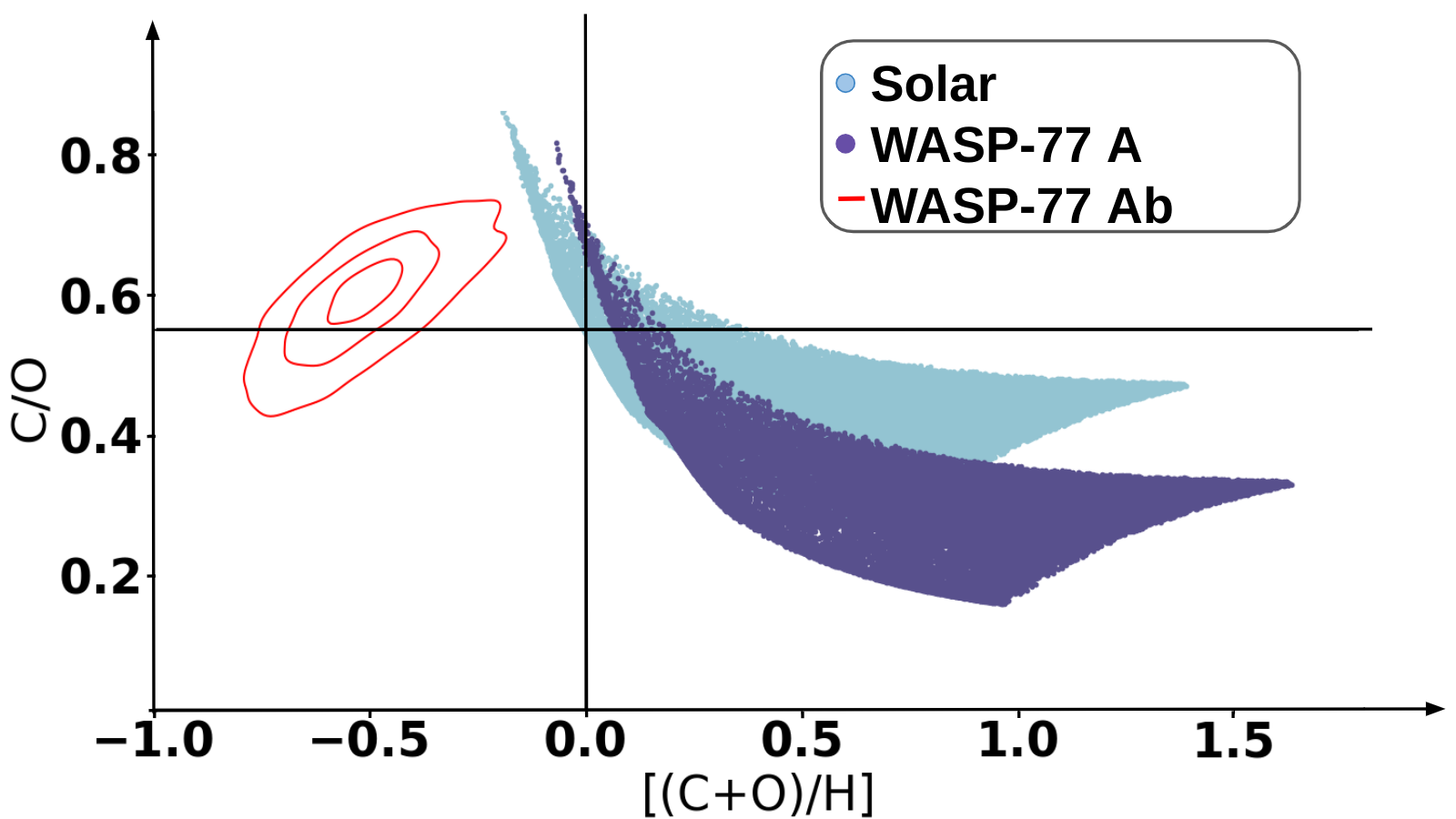}
                \caption{C/O ratio and \metallicity ~of a planet with a mass and orbital distance of \WASP ~formed in a disk of a solar composition (blue) and similar composition to WASP-77A (purple). Solid lines indicate the solar \metallicity ~and C/O ratio. The observed C/O ratio and \metallicity ~of \WASP ~and their contours are shown in red.
                }
                \label{Fig: init}
        \end{figure}
        
        However, by considering other scenarios as presented in the method section, it is possible to simulate planets with SimAb that end up having a similar composition to \WASP . We present two scenarios that demonstrate this below.\\

        \subsection{Scenario 1}
                        
        Assuming that the planet migrates via Type II migration all the way to its current location, we change the fraction of carbon that is present in the solid phase beyond the soot line, as explained in Section \ref{sec: method}. Figure~\ref{Fig: diskmig} shows that planets simulated by SimAb can acquire different C/O ratios and metallicities when different amounts of carbon is in the solid phase beyond the soot line.
        
        \begin{figure}[!htbp]
                \centering
                \includegraphics[height=4.5cm]{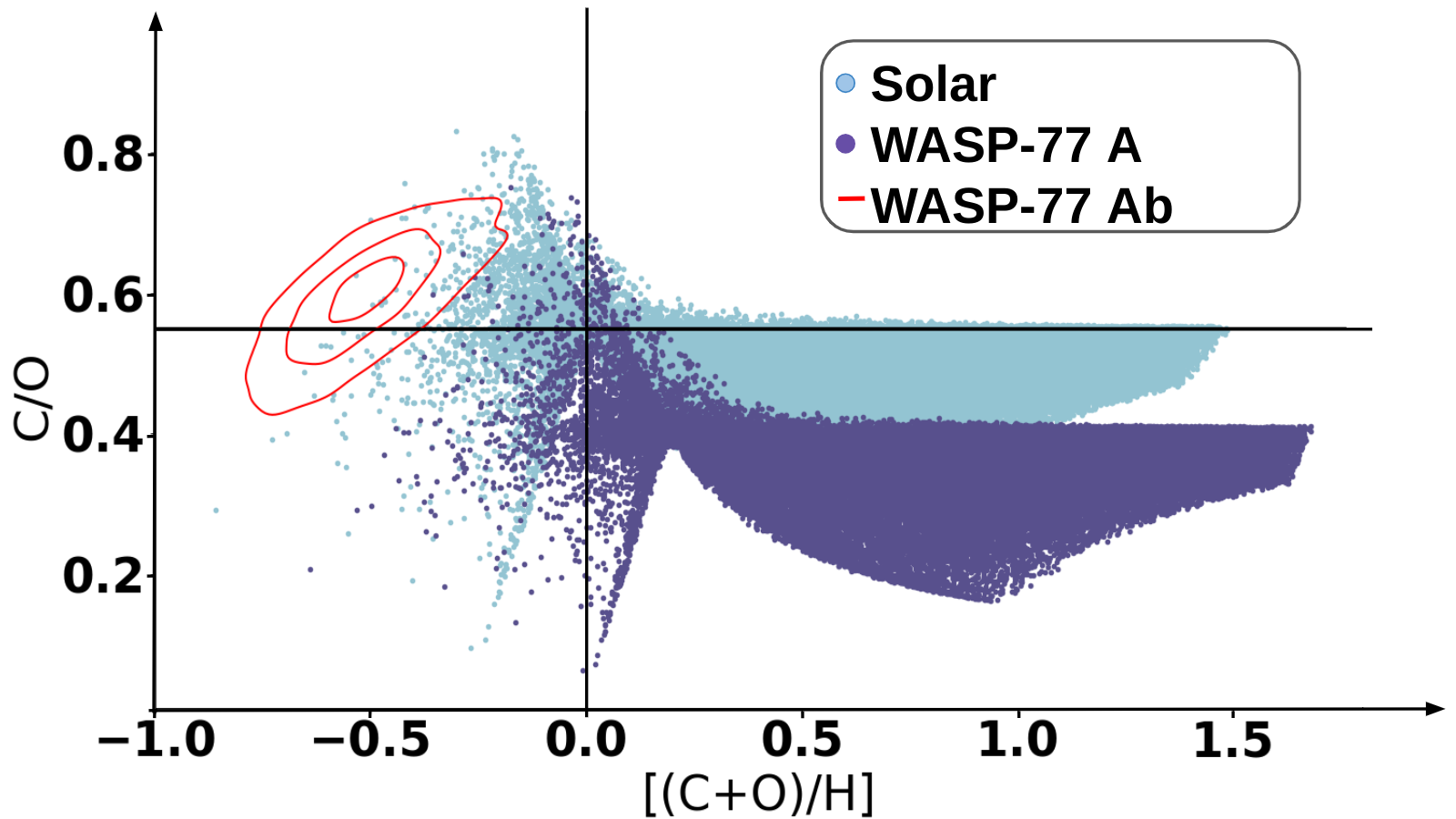}
                \caption{\textbf{Scenario 1}: C/O ratio and \metallicity ~distributions for planets with a mass and orbital distance similar to \WASP ~formed in a disk with a solar-like composition (blue) and a composition similar to the WASP-77A (purple). The observed C/O ratio and \metallicity ~of \WASP ~and their contours are shown in red. In this scenario, $f_{carbon}$ varies between 0 to 1.}
                \label{Fig: diskmig}
        \end{figure}

        By adding a soot line at 800K, SimAb produces planets with sub-stellar metallicities and sub-stellar C/O ratios. When assuming a higher carbon fraction in the solid phase at temperatures above 800K, planets would accrete a carbon-depleted gas. This allows the planets to achieve substellar metallicity and sub-stellar C/O ratios. This figure shows that there is more of an overlap between the simulated planets when including a varying carbon fraction at the soot line (see Fig \ref{Fig: diskmig}), as compared to a zero carbon fraction at the soot line (see Fig.~\ref{Fig: init}), where
there are more planets that have similar atmospheric composition to the observed composition of \WASP .

        Figure~\ref{Fig: retrieve4} shows the retrieved values for the initial orbital distance, dust fraction, planetesimal ratio, and carbon fraction at the soot line. The left panel showcases the scenario where the planet was formed in a disk with a solar composition, while in the right panel the disk is assumed to have a composition similar to WASP-77A. The results show that depending on the assumed composition of the disk where the planet was formed, the retrieved formation parameters differ.
        
        These plots show that given the precision reported for the atmospheric composition of \WASP , SimAb cannot significantly constrain the initial orbital distance of the protoplanet core. However, this suggests that the formation of a planet with a similar composition to the observed C/O ratio and \metallicity ~as \WASP ~is more likely when the planet initiates its migration beyond the CO$_2$ ice line if the disk had a solar-like composition. When assuming the disk composition is similar to the composition of WASP-77A, it is nearly impossible for the planet to have initiated its formation within the CO$_2$ ice line. 
        
        SimAb is able to provide an upper limit for the planetesimal ratio and dust grain fraction, which are found to be very low. This is expected due to the low observed \metallicity ~for \WASP. In both disk composition cases, the obtained amount of accreted planetesimal is found to be close to zero and the dust grains in the disk are shown to be the main source of the heavy elements in the atmosphere. The dust grain fraction is much lower when assuming the disk has a composition similar to WASP-77A, as compared to when the planet forms in a disk with solar composition (upper limits of $0.09$ and $0.29$ respectively). Additionally, SimAb predicts the planet must have formed in a disk where $48^{+18}_{-21}$ percent of the carbon is locked in the solid phase at the soot line for the disk with a WASP-77A-like composition and $55^{+23}_{-33}$ percent for the solar composition disk.

        \begin{figure*}[!htbp]
                \centering
                \includegraphics[height=7.6cm]{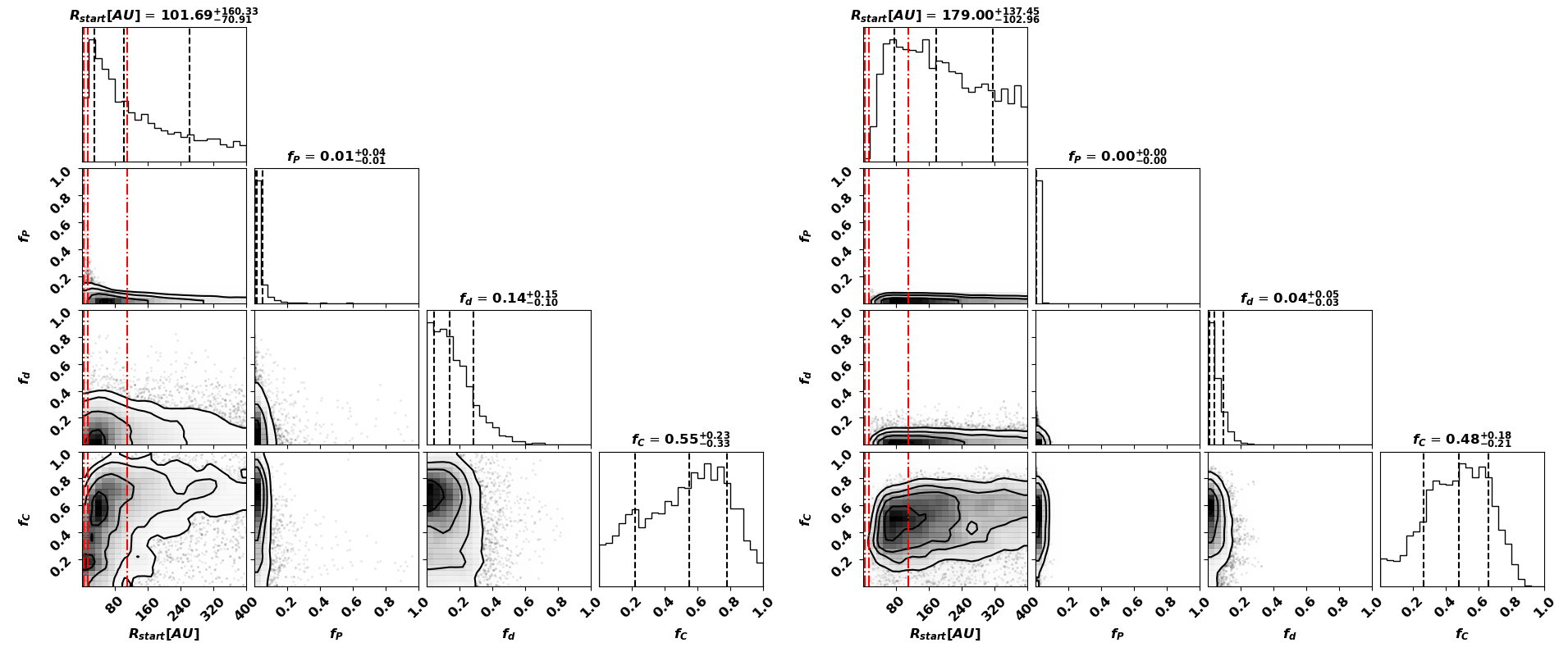}
                \caption{\textbf{Scenario 1:} Retrieved formation parameters of \WASP . Left panel shows the retrieved values for the case where the planet is formed in a disk with solar composition. Right panel shows the retrieved formation values for the planet when it forms in a disk with adjusted carbon and oxygen to represent the composition of WASP-77A. The red dashed lines show the three ice lines (i.e., water, CO$_2$, and CO) that are included in SimAb.}
                \label{Fig: retrieve4}
        \end{figure*}   

        In order to understand which region (with respect to the ice lines) is the more likely location for the planet to have initiated its formation, we looked at the four regions shown in Fig.~\ref{Fig: regions}, separately, assuming a disk composition similar to WASP-77A. Figure~\ref{Fig: region4} shows the overlap between the observed C/O ratio and \metallicity ~distribution (in black) and the predicted distributions based on the retrievals (in red) for each of these regions. There is a large overlap between the observed C/O ratio and \metallicity ~and the C/O ratio and \metallicity ~simulated by SimAb when the planet initiates its Type II migration beyond the CO$_2$ ice line and beyond the CO ice line. We emphasize that no {planet formation solution was found when assuming the planet initiated its Type II migration within the water ice line. 
        
        \begin{figure}[!htbp]
                \centering
                \includegraphics[height=6.5cm]{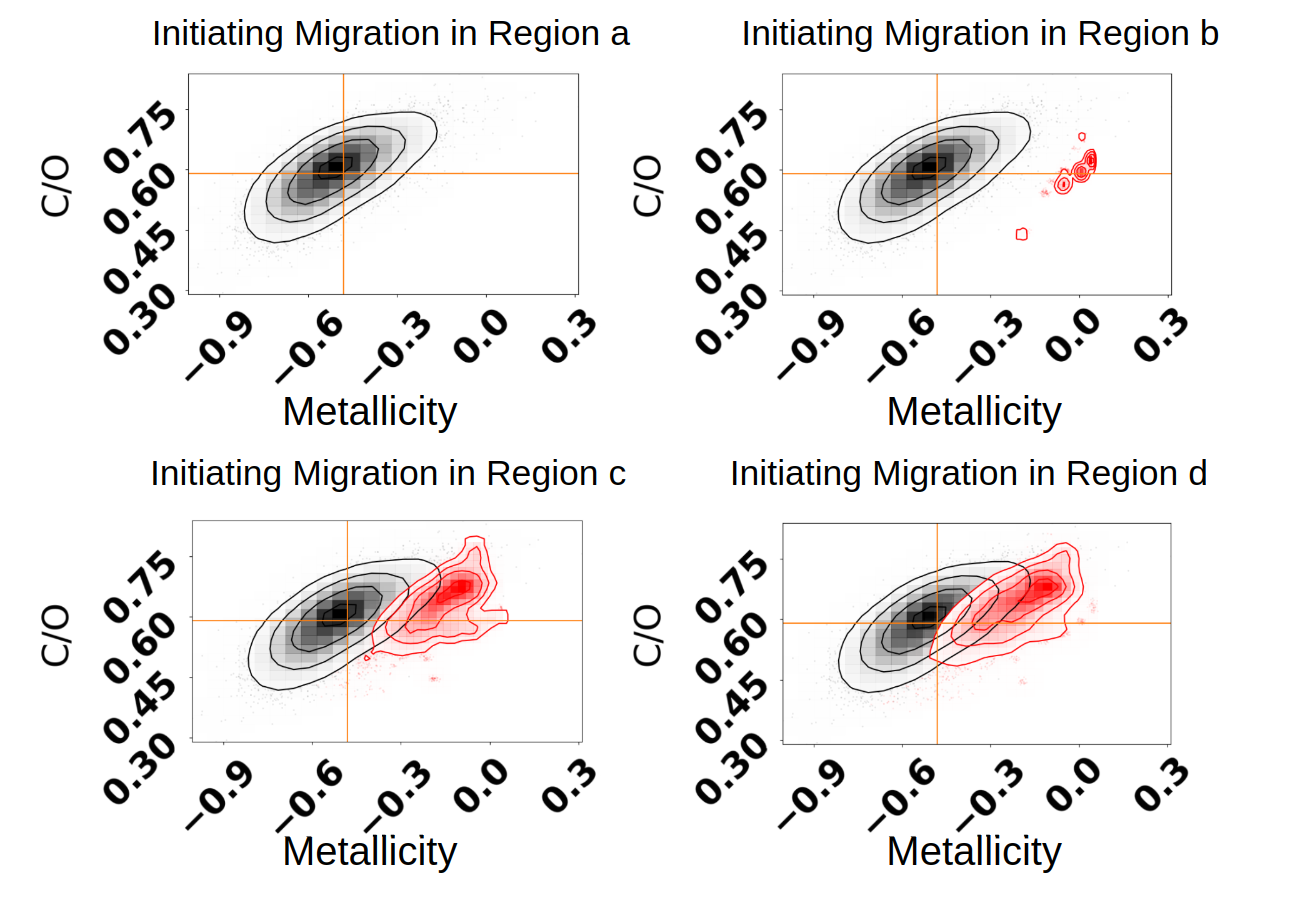}
                \caption{Comparison between the observed C/O ratio and \metallicity ~distribution (black) and the calculated distributions based on the first scenario formation retrieval, which assumes a disk with a composition similar to WASP-77A (red). The top left panel shows this distribution for planets that initiate their formation within the water ice line (no formation solution that matches this criteria was found). Top right panel shows this distribution for planets that initiate their formation between water and CO$_2$ ice line. Bottom left panel shows this distribution for planets that initiate their formation between CO$_2$ and CO ice line. Bottom right panel shows this distribution for planets that initiate their formation beyond the CO ice line.}
                \label{Fig: region4}
        \end{figure}
        
        \subsection{Scenario 2}

        Another possible scenario to form a planet with a similar composition as \WASP ~is for the planet to end its Type II migration beyond its current location. In this case, the planet would migrate to its current position after formation is finished without significant change to its composition. Figure~\ref{Fig: ndmig} shows planets with the same mass as WASP-77Ab ending their formation somewhere beyond the current location of \WASP . In this scenario, the carbon fraction at the soot line varies between 0 to 1. Planets that are fully formed beyond the CO ice line would have the same C/O ratio as their host star with various metallicities. This is seen as a line at C/O ratio $\approx$ 0.4 for the purple distribution and a line at C/O ratio $\approx$ 0.6 for the light blue distribution. This scenario does not explain the current orbital distance of the planet.

        \begin{figure}[!htbp]
                \centering
                \includegraphics[height=4.5cm]{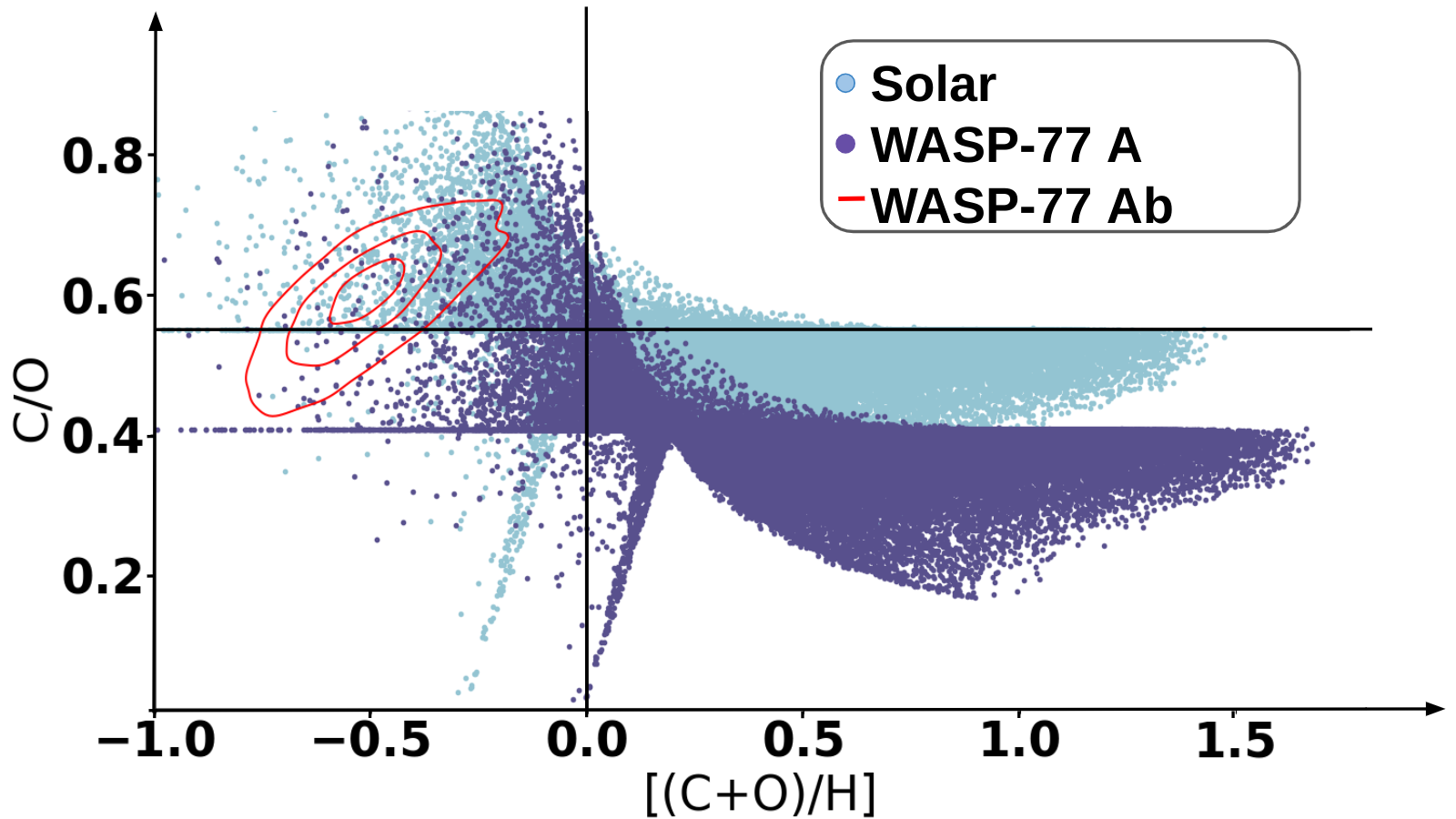}
                \caption{\textbf{Scenario 2}: C/O ratio and \metallicity ~distribution for planets with mass and final orbital distance similar to values seen for \WASP, either formed in a disk with solar composition (blue) and a composition similar to WASP-77A (purple). Observed C/O ratio and \metallicity ~of \WASP ~and their contours are shown in red. These planets end their migration somewhere beyond the current position of \WASP . The $f_{carbon}$ values for these simulations varies between 0 and 1.}
                \label{Fig: ndmig}
        \end{figure}
        
        Figure~\ref{Fig: retriev5} shows the retrieved values for this case including the final orbital distance to where the planet migrates and the migration distance during its mass accretion. The initial orbital distance was then derived using Eq.~(\ref{eq: distance}). This plot shows that regardless of the disk composition, there should be very little to no planetesimal accretion in order for the formed planet to have a similar composition as is observed for \WASP. On the other hand, the dust grain fraction obtained is found to be higher for planets forming in a disk with solar composition, with an upper limit of $0.32$ for a disk with solar composition versus an upper limit of $0.15$ for a disk with a composition similar to WASP-77A. The carbon fraction at the soot line is retrieved to be similar for both disk compositions, with $0.61^{+0.26}_{-0.38}$ for a disk with solar compositing versus $0.62^{+0.20}_{-0.32}$ for a disk with a composition similar to WASP-77A. For a planet forming in a disk with similar composition to WASP-77A, the final orbital distance is mainly constrained within the CO ice line and the initial orbital distance is mostly bound beyond the CO$_2$ ice line. However, for planets forming in a solar-like disk, the initial orbital distance retrieved is more likely to be anywhere beyond the CO$_2$ ice line and there is no constraint on the migration endpoint.
        
        To consider a more probable location for the formation of \WASP, ~we looked at the case where the planet is formed in a disk with a composition similar to that of WASP-77A. We considered four cases where the planet migration endpoint is in one of the four regions shown in Fig.~\ref{Fig: regions}. Figure~\ref{Fig: region5} shows the overlap between the observed CO ratio and \metallicity ~reported by \citet{Line_2021}, in black, as well as the C/O ratio and \metallicity ~that is produced by using the retrieved parameters as SimAb input. These figures show that there is an overlap between the two distributions when assuming the planet is formed within the CO ice line. Planets that are formed beyond the CO ice line may acquire similar \metallicity ~as the observed values for \WASP , however, their C/O ratio is the same as the host star, WASP-77A, and much lower than the observed C/O ratio of \WASP . The Bayesian evidence of these models is presented in Table (\ref{tab: sampels}).

        \begin{figure*}[!htbp]
                \centering
                \includegraphics[height=9cm]{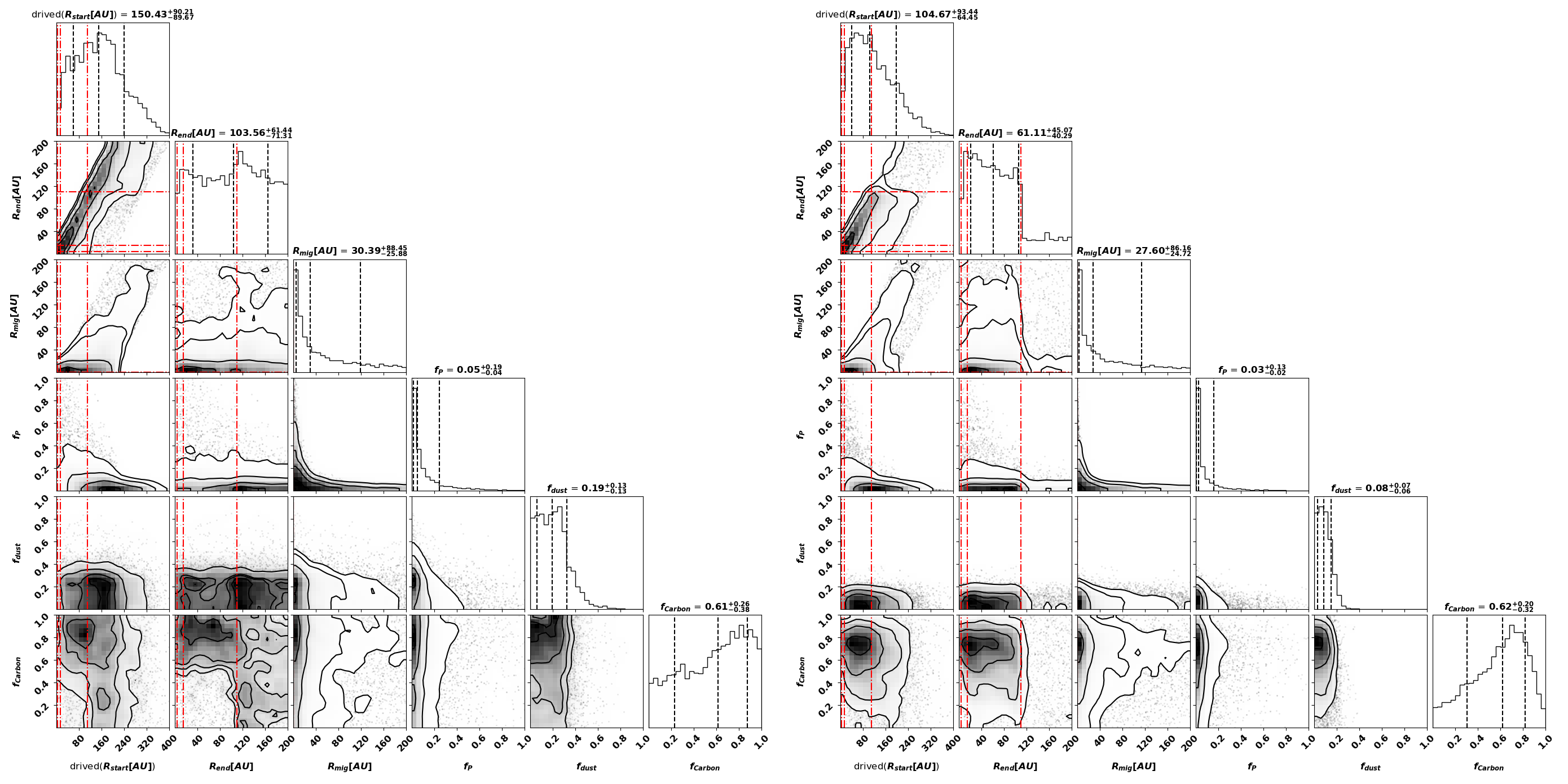}
                \caption{\textbf{Scenario 2:} Retrieved formation parameters of \WASP ~for the second scenario presented in Section \ref{sec2: formation scenarios}. Left panel shows the retrieved values for the case where the planet is formed in a disk with a solar composition. Right panel shows the retrieved formation values for the planet when it forms in a disk with adjusted carbon and oxygen to represent the composition of WASP-77A. The red dashed lines show the three ice lines (i.e., water, CO$_2$, and CO) that are included in SimAb. $R_{start}$ in these plots is a derived parameter using Eq.~(\ref{eq: distance}).}
                \label{Fig: retriev5}
        \end{figure*}

        \begin{figure}[!htbp]
                \centering
                \includegraphics[height=6.5cm]{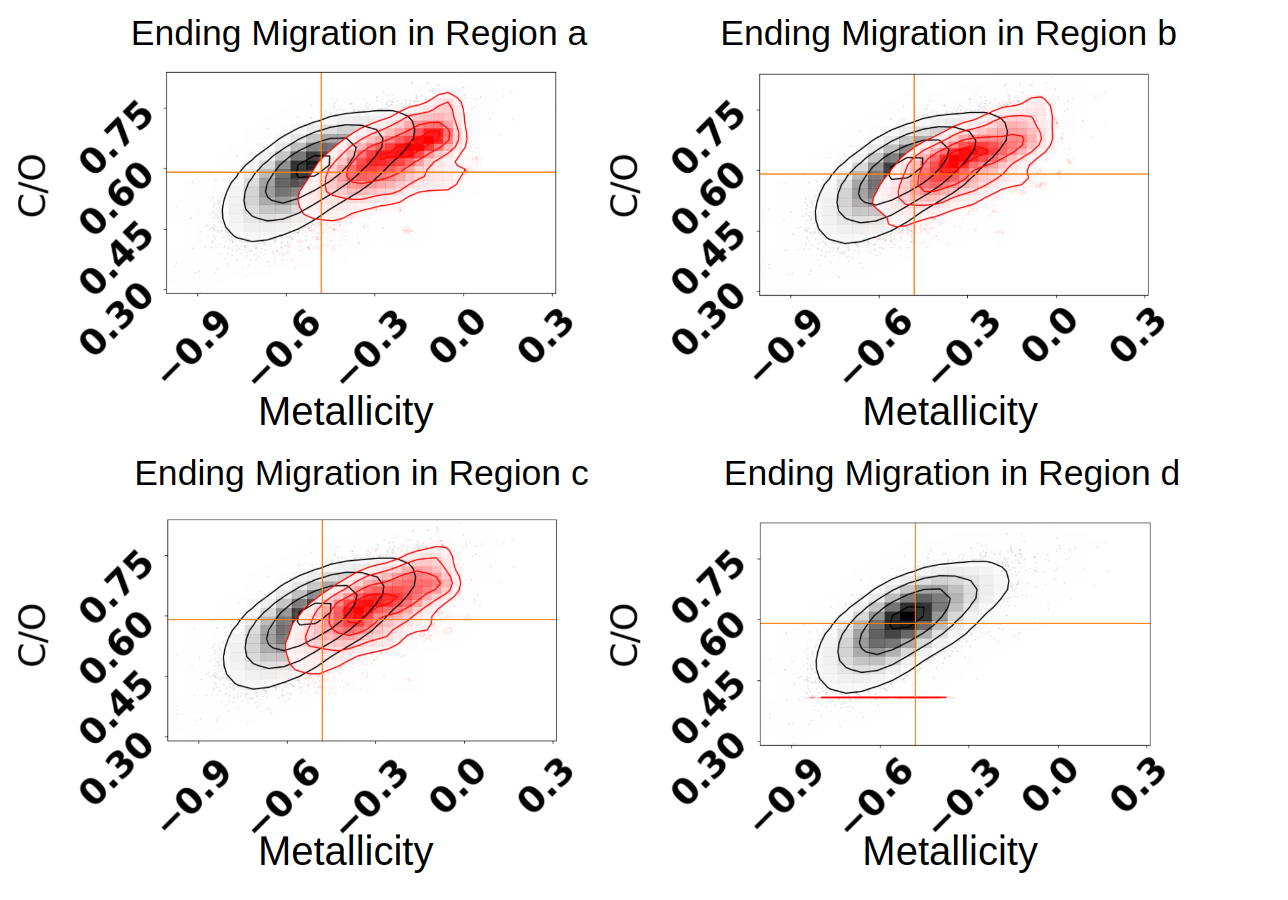}
                \caption{Comparison of the observed C/O ratio and \metallicity ~distribution (black) to that of the planets that end their Type II migration (red). Each panel shows this distribution for planets that end their Type II migration in the indicated region on top of the panel. The assumptions are the same as in Fig.~\ref{Fig: region4} but for \textbf{Scenario 2}.}
                \label{Fig: region5}
        \end{figure}

        \begin{table*}[ht!]
                \begin{center}
                        \begin{tabular}{ |c|c|c|c|c| } 
                                \hline
                                Scenario & Disk composition & Migration initial point [AU] & Migration endpoint [AU] & ln (Global Bayesian evidence) \\
                                \hline
                                
                                Scenario 1-all&Solar&all regions (400.0 - 0.024)&0.024& $-1.27$\\
                                Scenario 1-all&WASP-77A&all regions (400.0 - 0.024)&0.024& $-2.71$\\   
                                Scenario 1-a& WASP-77A&region a (3.63 - 0.024)&0.024& -\\   
                                Scenario 1-b&WASP-77A&region b (15.00 - 3.62)&0.024& $-7.71$\\  
                                Scenario 1-c&WASP-77A&region c (110.11 - 14.99)&0.024& $-3.36$\\  
                                Scenario 1-d&WASP-77A&region d (400.0 - 110.10)&0.024& $-2.09$\\
                                \hline
                                Scenario 2-all&Solar&all regions (400.0 - 0.025)&all regions (200.0 - 0.024)& $-0.39$\\   
                                Scenario 2-all&WASP-77A&all regions (400.0 - 0.025)&all regions (200.0 - 0.024)& $-1.29$\\     
                                Scenario 2-a&WASP-77A&all regions (203.63 - 0.025)&region a (3.63 - 0.024)& $-2.33$\\     
                                Scenario 2-b&WASP-77A&regions b, c, d (215.0 - 3.621)&region b (15.00 - 3.62)& $-0.87$\\     
                                Scenario 2-c&WASP-77A&regions c, d (310.11 - 14.991)&region c (110.11 - 14.99)& $-0.90$\\     
                                Scenario 2-d&WASP-77A&region d (400.0 - 110.101)&region d (200.0 - 110.10)& $-2.51$\\     
                                \hline
                                
                        \end{tabular}
                        \caption{Logarithmic global Bayesian evidence for each scenario}
                        \label{tab: sampels}
                \end{center}
                
        \end{table*}

\section{Discussion}
\label{sec: discuss}

\subsection{Formation in disks with different atomic abundances}
\label{sec1: dis-composition}

In this study, we look at the formation of \WASP ~assuming two different compositions for the disk where the planet was formed. As described in \citet{Reggiani_2022}, if we assume the planet was formed in a disk with solar atomic abundances, the observed C/O ratio of the planet is similar to its natal disk. However, assuming the composition of the disk is similar to the composition of WASP-77A leads to the conclusion that the observed C/O ratio of the planet  is instead higher than the disk's C/O ratio. In both cases, the observed \metallicity ~of the planet is lower than the assumed \metallicity ~of the disk. This implies that the formation scenario of WASP-77Ab would be different based on the original disk composition.

The Bayesian evidence for planet formation in a disk with solar composition is higher than that in a disk with a composition similar to WASP-77A. However, considering that WASP-77A has a different composition compared to that of the Sun, it is more realistic to assume \WASP ~was formed in a disk that is similar to the composition of its host star. By investigating the formation of \WASP ~in disks with both solar composition and with the WASP-77A composition, we show that the disk composition has an influence on our conclusions regarding {planet formation.
The main differences in the retrieved formation parameters when assuming two different compositions for the disk are seen in: the initial orbital distance where the planet initiates its Type II migration, final orbital distance where the planet ends its Type II migration, dust grain fraction, and the carbon fraction at the soot line.

The carbon fraction at the soot line impacts the C/O ratio in the disk within the water ice line as well as the carbon and oxygen abundance in the gas phase. A higher carbon fraction at the soot line results in a lower carbon abundance in the gas phase beyond the soot line and a lower oxygen abundance in the disk beyond the water ice line. 

On the other hand, the orbital distance where the planet initiates its Type II migration (in the first scenario) and the migration endpoint (in the second scenario) defines the regions where the planet accretes its atmosphere and the composition of the atmosphere. The results of this study show the migration distance is very similar in both disks with different compositions. The similarities among the migration distances could be related to our assumption on planetesimal accretion. We discuss this aspect in Section \ref{sec1: dis-limits}.

\subsection{Comparing the two scenarios}
\label{sec1: dis-scenarios}
In this section, we focus on the two scenarios we assumed when the planet was formed in a disk with a similar composition as WASP-77A. Comparing the Bayesian evidence of both scenarios, there is higher evidence for \WASP ~to have migrated to its current position through disk-free migration. Observations of the planet's eccentricity and inclination may provide evidence in support of this theory. Given the lack of confirmed companions in the outer orbits, the other possible candidate that may have caused a gravitational kick and led to the disk-free migration of the planet to its current orbit is WASP-77B ~\citep{Desidera_2006, Evans_2018}.

In both scenarios, the retrieved planetesimal fraction and the dust grain ratio are very similar and close to zero. However, this cannot put any constraints on the dust-to-gas ratio in the disk, unless further assumptions are made on the solid size distribution in the disk. The initial orbital distance that is retrieved in the first scenario and derived in the second scenario also shows some similarities. When comparing the first scenario to the second scenario, it is important to keep in mind that in the second scenario, the initial orbital distance is not a retrieved parameter but, rather, a derived parameter, which has a linear correlation with the migration distance and the migration endpoint through Eq.~(\ref{eq: distance}). This dependency shapes the distribution.

In both scenarios, as shown in Fig.~\ref{Fig: retrieve4} and Fig.~\ref{Fig: retriev5}, it is more likely for the planet to have initiated its migration beyond the CO$_2$ ice line. However, in the first scenario, the planet accretes material from everywhere in the disk within its migration initial orbital distance. This means that the planet is forced to accrete material from within the water ice line. This is a region where the oxygen abundance in the gas phase becomes very high compared to hydrogen and its accretion would result in the formation of planets with relatively higher \metallicity ~compared to the second scenario (see Figs.~\ref{Fig: region4} and \ref{Fig: region5}). Therefore, in the second scenario, a smaller initial orbital distance is possible. Additionally, the retrieval of the migration distance in the second scenario shows that the planet goes through shorter migration distances compared to the first scenario, avoiding atmospheric accretion from within the water ice line. Figure \ref{Fig: region5} shows that on average, planets that end their migration within the water ice line exhibit higher \metallicity . Moreover, the migration distance impacts the amount of planetesimals accreted onto the planet. Hence, a higher planetesimal fraction is possible in the second scenario, as compared to the first scenario.

Even though the carbon fraction at the soot line is seen to peak at a higher value in the second scenario compared to the first scenario, the two values are consistent within one sigma. In both scenarios, the retrieved values for the carbon fraction at the soot line peaks at a higher value compared to what is observed in the solar system \citep{Lodders_2009,Asplund_2009,Lee_2010}. Our results show consistency with the studies on the carbon fraction in the interstellar medium (ISM) \citep{Gail_2017,jones_2011}.

\subsection{Comparisons with previous studies}
\label{sec1: dis-compare}
Our results agree with the conclusion reported in \citet{Reggiani_2022}, where the authors claim the planet should have formed in the disk beyond the water ice line, regardless of the composition of the natal disk. Our study puts further constraints on the location where the planet was formed, based on the chosen formation model and the disk composition (discussed in Section \ref{sec1: dis-scenarios}). However, our results are not in agreement with what is suggested in \citet{Line_2021}, where the authors claim the low C/O ratio is achievable if the planet was formed within the major ice lines in a gas with a low carbon abundance. Even though formation within the major ice lines (assuming the majority of the carbon is in the solid phase) can explain the low C/O ratio, it cannot explain the depletion in oxygen.

Furthermore \citet{Bitsch_2022} studied the formation of \WASP ~and $\tau$ Bo{\"o}tis. In their model, the authors assume a solar-like composition for the disk, while allowing for changes to the C/O ratio of the disk due to pebble drift and evaporation. Their results suggest that assuming a solar-like composition for the disk, in order to form a planet with the observed carbon and oxygen abundance characterizing \WASP,~the planet must have completely formed beyond the CO$_2$ ice line and had been kicked inward. This is in agreement with our results from the second scenario, assuming a solar composition for the disk where \WASP~was formed. However, our results also allow for the planet to have migrated past the water ice line. This difference in our results could be a result of allowing for different carbon fractions at the soot line, which causes further depletion in carbon and oxygen from the gas phase outside the water ice line in comparison with the disk composition in \citet{Bitsch_2022}.

\subsection{Limitations of the model}
\label{sec1: dis-limits}

To calculate the atomic abundances of the disk in the gas phase and the solid phase, we used a simple procedure (explained in Paper I). Furthermore, we assumed a steady-state disk where the planet's formation or passing time would not impact the composition, the temperature, or the density of the disk. This results in the disk solid-to-gas atomic ratio remaining constant between the ice lines. By considering a more detailed atomic gradient between the ice lines, the retrieved migration parameters may be constrained with a higher level of precision.

The migration of the planet, however, impacts the mass that is accreted in each region. As a consequence, this can change the composition of the accreted atmosphere. The accreted mass in each region is also dependent on the mass of the protoplanet at each step. Even though, in our first study, we showed that the atmospheric composition of the gas giants is not correlated with their initial core mass, it can impact the values that are retrieved in this study -- but not the overall conclusion regarding the formation of the planet. In other words, the model will accurately represent which region the planet has formed in, but not necessarily the exact position at each step.

Additionally, our assumptions regarding accreting planetesimals impact the results with respect to the migration of the planet. We assume the planetesimals are accreted throughout the formation process and the amount of accretion is dependent on the $f_p$ and the distance that the planet travels as it migrates. This directly impacts the derived migration distance. This impact is even more evident for planets with super-solar metallicities and care needs to be taken when studying such planets. In the current study, this impact is negligible as the planetesimal ratio is found to be very low in all scenarios.

By adjusting the oxygen and carbon abundance to the values observed for WASP-77A, while not adjusting the other elements results in an excess of oxygen in the solid phase after the water ice line. Considering that the abundance of oxygen is between one to two orders of magnitude higher than the other refractory elements, this effect can be ignored. However, this effect should be studied if looking at other elements in the planetary atmosphere.

In the work of \citet{Line_2021}, the impact of rain out on their observation is discussed. Rain out in the atmosphere removes part of the oxygen from the observable layers of the atmosphere. In this case, the observed oxygen abundance will be lower than the total atmospheric oxygen abundance. This indicates a lower atmospheric C/O ratio and higher \metallicity ~compared to the observed values; therefore, assuming rain out in the atmosphere will impact the retrieved formation parameters. However, including this scenario is beyond the scope of this study.

As is shown in Figs. 7-9 of Paper I, planets with C/O ratios and metallicities close to those of their host stars will, in principle, need more precise information on their C/O ratio and metallicity in order to constrain their formation parameters. Even though the observational precision in \citet{Line_2021} are among the highest for ground-based observations, the precision is still too low to be able to put further constraints on the formation parameters. However, we have to realize that greater accuracy may not be enough to lift the inherent degeneracies in our current understanding of {planet formation.

The processes of {planet formation and evolution are complex and can have a sizable impact on the composition of their respective planets. In this study, we consider the roles of major players in {planet formation, however, the inclusion of other processes may affect the results \citep{Molliere_2020}. Thus, it is important to compare such studies to ours to get a better understanding of the processes that were involved in the formation of \WASP ~and to obtain a more complete picture of its formation history.

\section{Conclusion}
\label{sec: conclusion}

    In this work, we retrieved the formation parameters of \WASP ~based on the observed C/O ratio and the \metallicity ~reported by \citet{Line_2021}. We used the SimAb {planet formation simulation and MultiNest to study the formation parameters space and find the most likely combination to form a planet with a similar C/O ratio and \metallicity ~as the observed values for \WASP . Our results show that it is less likely for the planet to have initiated its formation within the water ice line. Considering that the composition of WASP-77A should be a better representation of the disk composition where \WASP ~was formed, as compared to a solar composition, it is likely that the planet initiated its migration beyond the CO$_2$ ice line. The more likely scenario for the formation of the planet is that the planet formed somewhere beyond its current location and was moved inward via disk-free migration. In this case, the planet is expected to have accreted the majority of its material between the water ice line and the CO ice line. Assuming a solar composition for the disk also allows for the accretion of the atmosphere beyond the CO ice line. Our study shows that scenarios in which a significant amount of the carbon stays in the solid phase beyond the soot line are more favorable in comparison to scenarios that replicate the carbon fraction in the solar system. Depending on the formation scenario, this value could vary between an average of $48 \%$ to $62 \%,$ which is in agreement with ISM observations. It is important to note that the carbon fraction at the soot line is not tightly constrained by the observations, so the uncertainties on this value are large. Given its very low \metallicity, the planet could not have accreted many planetesimals during its process of atmospheric accretion. This is made evident by the very low value obtained for the planetesimal ratio.

\section*{Acknowledgements}

      J.M.D acknowledges support from the Amsterdam Academic Alliance (AAA) Program, and the European Research Council (ERC) European Union’s Horizon 2020 research and innovation program (grant agreement no. 679633; Exo-Atmos). This work is part of the research program VIDI New Frontiers in Exoplanetary Climatology with project number 614.001.601, which is (partly) financed by the Dutch Research Council (NWO). 

      We would like to thank the authors of \citet{Line_2021} for making their data publicly available which enabled us to do this research.

        \newpage
        \bibliographystyle{aa}
        \bibliography{reference}


\end{document}